\documentclass{revtex4}
\usepackage{epsfig}
\begin{document}
\title{Hysteresis in an Ising model with mobile bonds}
\author{D.~\v Capeta and D.K.~Sunko\thanks{email: dks\@phy.hr}}
\affiliation{
Department of Physics,\\Faculty of Science,\\
University of Zagreb,\\
Bijeni\v cka cesta 32,\\
HR-10000 Zagreb, Croatia.}
%\date{}

\begin{abstract}
Hysteresis is studied in a disordered Ising model in which diffusion of
antiferromagnetic bonds is allowed in addition to spin flips. Saturation
behavior changes to a figure-eight loop when diffusion is introduced. The
upper and lower fields delimiting the figure-eight are determined by the
Hamiltonian, while its surface and the crossing point depend on the
temperature and details of the dynamics. The main avalanche is associated
with the disappearance of hidden order. Some experimental observations of
figure-eight anomalies are discussed. It is argued they are a signal of a
transient rearrangement of domain couplings, characteristic of amorphous
and/or magnetically soft samples, and similar to evolution of kinetic
glasses.
\end{abstract}

%\pacs{02.70.Ns, 05.70.Ce, 31.15.-p}

\maketitle
\section{Introduction}

Magnetic hysteresis is a classic example of metastability at
experimentally easily accessible time scales. Its ready appearance under
simple conditions makes it not only technologically important, but also a
theoretical challenge. In addition to macroscopic approaches, relying on
multiply-valued equations of state, there has been considerable work done
over the past decade in Monte Carlo simulations of lattice systems. The
high reproducibility of hysteresis curves, and the fact that among their
many observed shapes most are fairly simple, imply that hysteretic delay
involves processes similar to ordinary thermalization, which can be
modeled by a simple random walk. It is thus reasonable to ask, which
features need to be added to the walker in order to reproduce some
particular aspect of the experimental observations.

The Ising model, in the widest sense, has evolved as an important test-bed
for this kind of question. Its spins are naturally associated with
magnetic domains, and can be coupled by different physically motivated
interactions. It turns out that random fields with cluster updates give
rise to Barkhausen noise~\cite{Sethna93}, local spin inversion symmetry is
needed for memory effects~\cite{Katzgraber02}, random couplings produce
multicycles at low temperatures~\cite{Deutsch03}, while long-range and
anisotropic interactions give domain patterns very similar to those in
some real thin films~\cite{Iglesias02}.

The present work reports one more result of this kind: allow local
diffusion of antiferromagnetic (AF) couplings, and the saturation part of
the hysteresis curve will change, from the usual horizontal line to a
figure-eight form. Such anomalies are sometimes observed in experiments on
thin films and tapes, when the coercive fields are of the order of 100~Oe.
We are aware of only one paper, however, where the authors have
specifically commented upon them~\cite{dosSantos95}. It is known from a
diffraction experiment, albeit with domain structure carefully controlled
by micromachining, that spatial evolution of AF couplings can occur under
driving by a field~\cite{Chesnel02}. We suggest here that a physically
similar, but random, mechanism may be responsible for figure-eight shapes
in saturation curves, in the usual situation, where sample domain
structure is left to chance.

\section{The model}

We use the same model as before~\cite{Lazic01}. Take the two-dimensional
short-range bond-disordered Ising model, or $\pm J$-Edwards-Anderson
(EA) model~\cite{Edwards75}
\begin{equation}
H=-\sum_{<i,j>}J_{ij}S_iS_j-B\sum_iS_i
\label{model}
\end{equation}
where $J_{ij}=\pm J$ and $S_i=\pm 1$. The model is subject to
Glauber~\cite{Glauber63} dynamics: spin is flipped at random, and the move
is accepted if the criterion
$$
\frac{1}{1+\exp(\Delta/T)}>c
$$
is satisfied, where $\Delta$ is the energy balance of the trial move, $T$ the
temperature, and $c$ a uniformly distributed random number, $0<c<1$. Between
each two spin trials there is a bond trial: a positive and negative bond
impinging on the same site, and chosen at random, are allowed to exchange
places, subject to the same criterion as the spin trials. The concentration
$p_{AF}$ of antiferromagnetic (AF) bonds is taken to be 50\%, unless
otherwise noted.

This dynamics causes a significant annealing of the sample, and after a
transient of $\sim 100$ updates per site it enters a long-lived state
characterized by glass-like dynamics, with a second, much longer relaxation
time~\cite{Lazic01}. The `glassiness' of the metastable state is not
topological: the underlying bond distribution, viewed at any instant in time,
has low frustration, so much so that it can be mapped onto a disordered
ferromagnet~\cite{Hartmann03}, with a finite transition
temperature~\cite{Merz02}. However, when the bonds move, the spins are
prevented from settling down into any given ordered arrangement. In this
sense the situation is more reminiscent of the `geometric' frustration of
vitreous liquids, than of spin glasses: configurations are locally relaxed,
but the spin correlation decays over time even at low temperature. As long as
the number of F and AF bonds is kept equal, the equilibrium zero-field
magnetization is zero.

The equivalent disordered ferromagnet is obtained by local gauge
transformations which change the sign of bonds whenever the plaquette is not
frustrated. The disorder in the ferromagnet comes from the residual AF bonds,
related to the frustrated plaquettes, which are relatively few, so a net
magnetization can appear, revealing the formerly hidden order. (This is
different from a dilute ferromagnet, where some bonds have been assigned
$J=0$.) The degree of frustration, hence the order associated with the hidden
ferromagnet, can change as a function of external field, because of bond
diffusion. This is specific to the present model. In the usual situation
without bond updates, the frustration is fixed for a given sample, and when
$p_{AF}=50$\%, the equivalent ferromagnet has no phase transition at finite
temperature, and neither has the spin glass.

\section{Main result}

\begin{figure}
\center{\epsfig{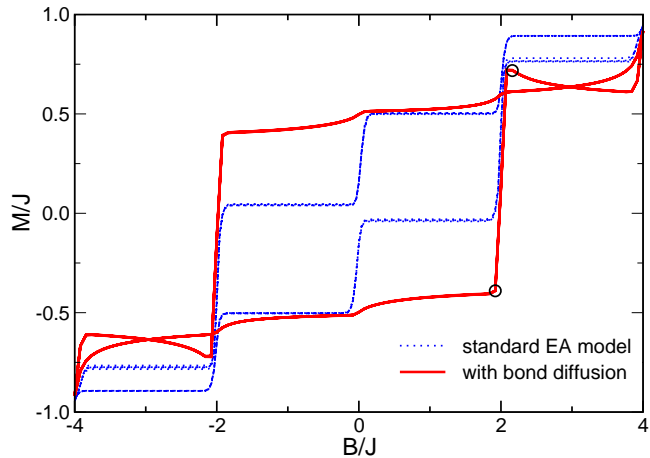}}
\caption{Hysteresis in the model~(\ref{model}), with (full curve) and without
(dotted curve) bond updates. Time spent at each field is ten updates per
site. The initial parts of the loops have been erased for clarity. Note that
the half-width of the main loop is the same as the range of each figure
eight, namely $2J$. Open circles: see Fig.~\ref{figconfig}.}
\label{figmain}
\end{figure}
The model is subjected to a periodic external field, starting with $B=0$ and
increasing to $B=4J$, then reducing the field to $B=-4J$, and finally closing
the curve. At each point in the field, we take 10~updates per site (meaning
10~spin and 10~bond updates in alternation). The dependence on this number
(sweep rate) will be discussed in the next section. Two field steps were
used, $\Delta B=0.04J$ and $\Delta B=0.08J$, with no discernible difference
in the results. We note that for our two-dimensional square lattice, $B=4J$
is the limit at which saturation is enforced, at a magnetization determined
only by the temperature. The outcome is shown in Fig.~\ref{figmain}, compared
to the case without bond updates. The latter has been studied
before~\cite{Vogel01} for the 3D case. Its hysteresis curve consists of a set
of box-like regions, with constant magnetization alternating with large
avalanches. This is easily understood, because the system is essentially in
the clean limit. Namely, an individual Ising spin has only a finite, discrete
set of energy values available. As soon as the external field passes through
one of these values, all the respective spins are destabilized at the same
time, and there is nothing to stop the ensuing avalanche. This result
corresponds to low temperature ($T=0.1J$), on which we concentrate in the
greater part of the article. Note that with a discrete set of bond strengths,
the system is gapped, so the temperature $T=0.1J$ gives nearly the same
results as $T=0.01J$, except that the figure eight is slightly shifted, an
insignificant effect to the purposes of this article.

Bond updates are now introduced, alternating equally with spin updates. It is
no longer necessary to average over samples, because bond movement makes the
whole sample space available to a single walker, or in technical terms, it
induces self-averaging. This is so efficient that Fig.~\ref{figmain} simply
gives the raw data for five cycles, showing how the corresponding points fall
on the same curve, meaning the error bars are smaller than the line
thickness. The simulations were mainly on $100\times 100$ lattices, with
spot-checks to make sure that $50\times 50$ and $200\times 200$ gave the same
results. The curves were also independent of the random initial conditions,
given that field cycling began after the above-mentioned fast transient
($50$--$100$ MCS per site) died out.

Bond movement introduces two new features in Fig.~\ref{figmain}. First, the
two inner boxes (around the origin) open up into a single large one, with
avalanches at $B=\pm 2J$. This is similar to what happens when a random
static site field is added to the model without bond movement, or for a
Gaussian distribution of bond strengths. It can be explained by the
disappearance of spins intrinsically in an indifferent state, surrounded by
the same number of positive and negative bonds, which are responsible for the
avalanches at zero field. However, the reason for the disappearance is
somewhat different in the two cases. Introducing site disorder of some
characteristic magnitude $K$ will make the formerly indifferent spins
unstable at some new field values $B\approx\pm K\neq 0$, leaving few to
participate in the avalanche at $B=0$. In the present model, on the other
hand, bond movement induces significant relaxation of the system, meaning
most spins are in a locally minimal configuration, so again the number of
indifferent ones is low.

\begin{figure}
\center{\epsfig{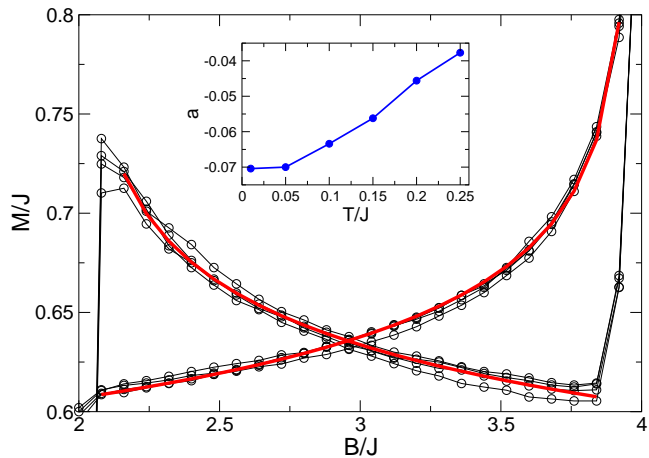}}
\caption{Magnified figure-eight anomaly of Fig.~\ref{figmain}, fitted (thick
curves) to $B^a$ with $a=-0.07$. The value of $a$ is the same in the
ascending and descending parts. Circles are raw data over four cycles. Inset:
dependence of the exponent on the temperature.}
\label{figpower}
\end{figure}
The second new feature is somewhat unexpected. The former small box at
$2J<B<4J$ is now a distinct figure-eight loop. Both the forward (increasing
$B$) and backward (decreasing $B$) parts of the magnetization curve have the
form of relaxation curves. They are given by power laws in $B$ with small
exponents, as shown in Fig.~\ref{figpower}. By contrast, the temporal
autocorrelation of the magnetization is a stretched exponential, as in other
similar models. Hence the $M(B)$ curves cannot be interpreted as a simple
temporal evolution, trivially due to the fact that $B$ changes before the
system has had time to equilibrate. The bond updates introduce a new
relaxation mechanism, with a scale comparable to the scale of the main cycle.
The exponent of $B$ becomes (absolutely) smaller as the temperature increases,
which reflects the figure-eight becoming thinner and flatter. The temperature
dependence of the exponent is given in the inset to the figure.

\begin{figure}
\center{\epsfig{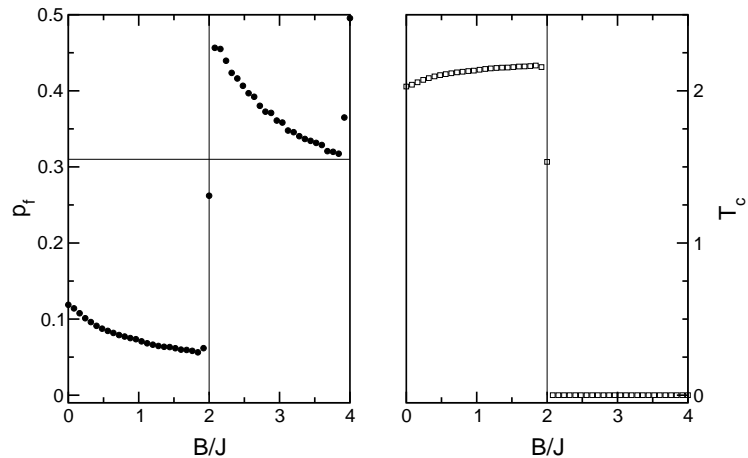}}
\caption{Left: concentration of frustrated plaquettes as a function of the
field, for the full curve in Fig.~\ref{figmain}. Horizontal line: critical
value $p_c=0.31$ at which $T_c=0$ for the equivalent disordered ferromagnet.
Right: transition temperature of the latter, obtained from the values in
the left panel according to Ref.~\cite{Merz02}.} \label{figfrustr}
\end{figure}
The main avalanche in this model is microscopically different than in models
with quenched disorder, because in the presence of bond updates, the
frustration of the lattice can vary as a function of the field. This is shown
in Figure~\ref{figfrustr}. In the region of the main loop, the equivalent
disordered ferromagnet has a finite transition temperature. The concentration
of frustrated plaquettes rises sharply with the main avalanche, and passes
through the critical value, above which the ferromagnet has $T_c=0$. In this
way the spin avalanche is connected with a ground-state transition in the
bond lattice. For $B<2J$, the instantaneous bond configuration can support an
ordered spin state at finite temperature; above this value, it cannot.

\begin{figure}
\center{\epsfig{file=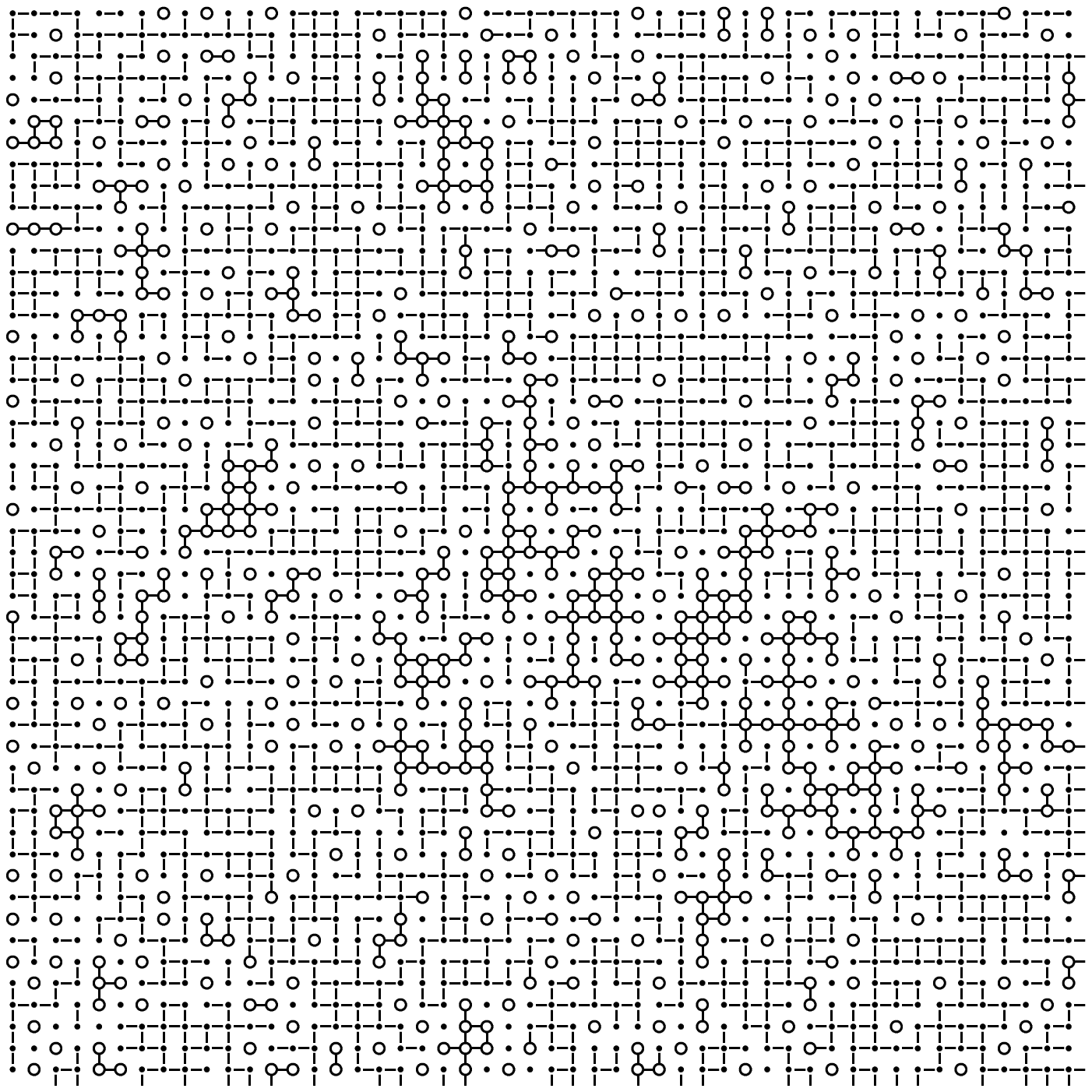,height=60mm}
\epsfig{file=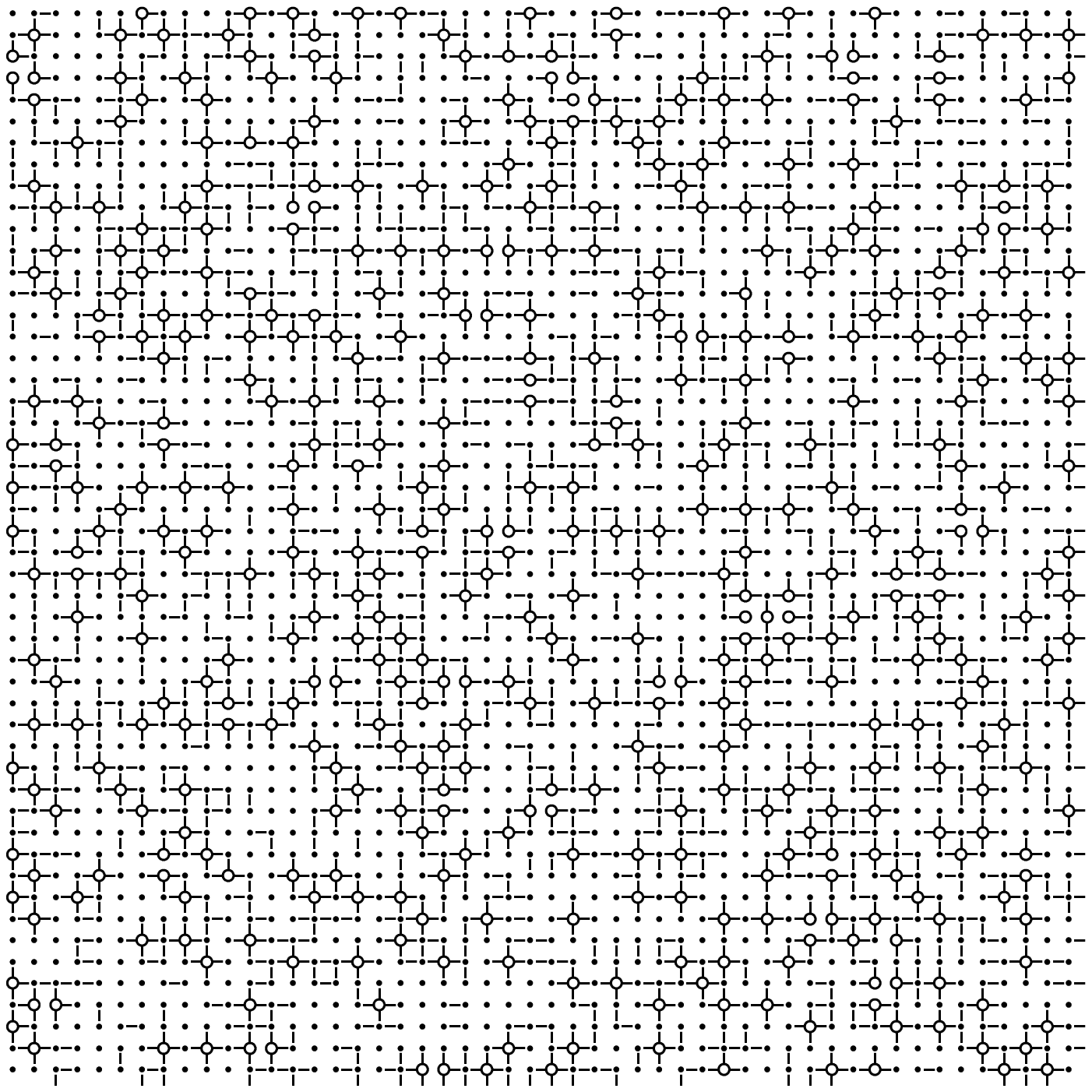,height=60mm}}
\caption{Configurations before and after the main avalanche. Left: 
ordered state, lower open circle in Fig.~\ref{figmain}. Only ferromagnetic
bonds are shown. Right: disordered state, upper open circle in
Fig.~\ref{figmain}. Only AF bonds are shown. Circles: down-spins. Dots:
up-spins.}
\label{figconfig}
\end{figure}
In Fig.~\ref{figconfig}, sample configurations are shown immediately
before and after the main avalanche. In the ordered state, there are
islands of spins with the same orientation (dot: up, circle: down),
connected by ferromagnetic bonds. In the disordered state, the islands of
down-spins (circles) have broken up, and each is preferentially connected
with up-spins by `crosses' of AF bonds.

\section{Phenomenology of the model}

In this section, we show how the basic curve in Fig.~\ref{figmain} changes
with various aspects of the model dynamics. There are three to which it
should be at least moderately robust: temperature, ratio of spin to bond
updates, and sweep rate.

\begin{figure}
\center{\epsfig{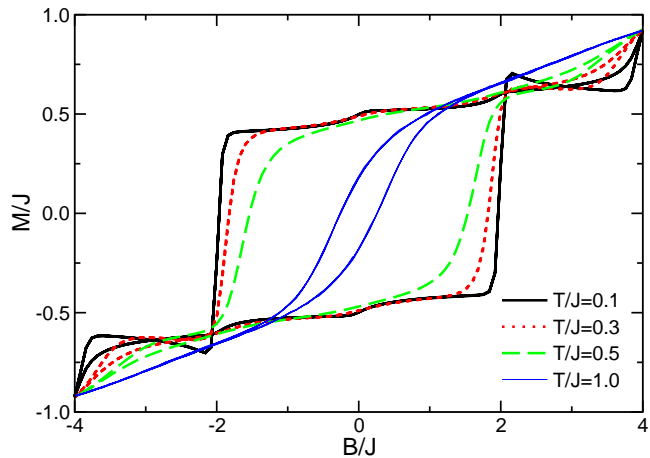}}
\caption{Temperature dependence of the hysteresis. Thick full line: $T=0.1J$
(Fig.~\ref{figmain}). Dotted line: $T=0.3J$. Broken line: $T=0.5J$.
Thin full line: $T=J$.}
\label{figtemp}
\end{figure}
In Figure~\ref{figtemp}, we observe that the figure-eight evolves with
temperature on the scale of $J$. At $T=0.5J$, the first maximum of the figure
eight has dropped below the reverse curve, so there is now an open narrow
loop at $B>2J$, which is still wider at the middle than at either end. The
narrowing of the main part $-2J<B<2J$ is significantly milder than in the
case without bond updates~\cite{Vogel01}, so that, overall, their
introduction makes the hysteresis less sensitive to temperature increase.

\begin{figure}
\center{\epsfig{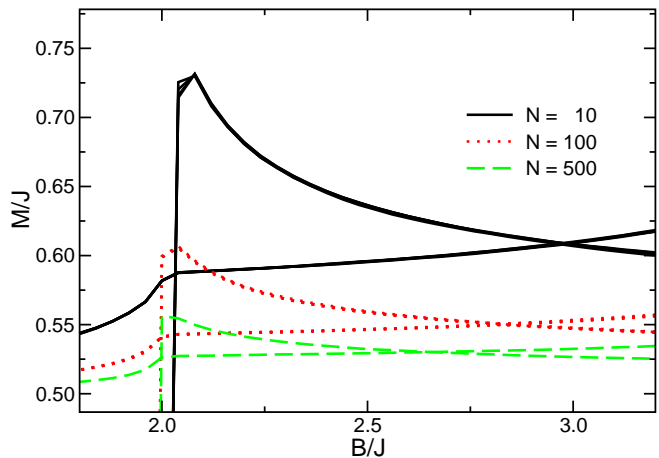}}
\caption{Dependence of the figure-eight feature on the time spent at
each field. Full line: 10~updates/site
(Fig.~\ref{figmain}). Dotted line: 100~updates. Broken line: 500~updates.
The field step is $\Delta B=0.04J$ in all three curves.}
\label{figupdates}
\end{figure}
In Figure~\ref{figupdates}, we show that the figure-eight feature is not a
transient effect due to too fast cycling. It visibly persists to up to
500~updates/site at each field, although it becomes thinner. We know that in
the model's two-time dynamics, the transition between the fast (transient)
and slow (metastable equilibrium) component occurs at 50--100 updates per
site, so the feature survives into the intrinsic `glassy' regime. Observed
figure-eight anomalies are indeed thinner (relative to the main loop) than
suggested by Fig.~\ref{figmain}. This agrees with the idea that
experimentally, magnetizations at each point of the loop are measured after
any fast transients have died out. We find, nevertheless, that the power-law
relaxation of $M$ with $B$ is found in all three curves in
Fig.~\ref{figupdates}, so it is not wrong to use fast cycling for model
investigations. Only the numerical value of the relaxation exponent changes,
decreasing with slower cycling. Increasing the number of updates at each
field also gets rid of the steps at $B=0$ in Fig.~\ref{figmain}, without
changing the shape and size of the main loop.

\begin{figure}
\center{\epsfig{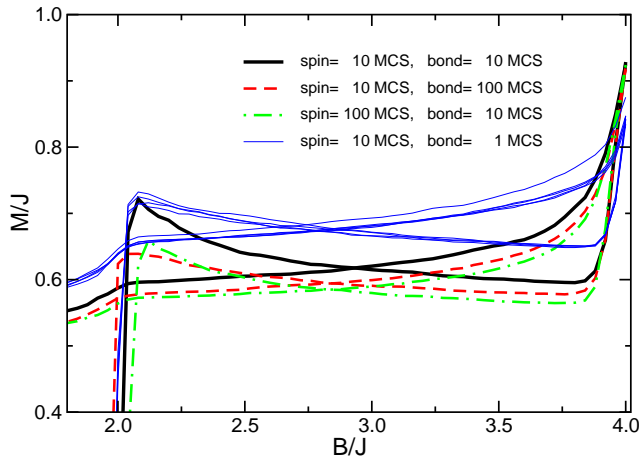}}
\caption{Dependence of the figure-eight feature on the ratio of spin to
bond updates, at $T=0.05J$. Raw data for five cycles are shown for each curve.
See the text for details.}
\label{figratio}
\end{figure}
In Figure~\ref{figratio}, we show perhaps the most unexpected outcome of
the phenomenological investigations. We vary the relative frequency of
spin and bond updates, guided by the idea that they are not equally common
in real materials. Based on past modelling experience, one would assume
that bond updates should be rarer, so it matters that the figure-eight is
not too sensitive to their rate. Indeed it is not, but it also turns out
that the system responds much the same whether it is the rate of spin or
bond updates that is decreased. Hysteresis curves in which they differ by
a factor of ten either way fall practically on top of each other. Bond
updates are therefore efficient in mutually equilibrating the spin and
bond subsystems. They are out of equilibrium with respect to the
asymptotic state, but not with respect to each other. Figure~\ref{figratio}
also shows a dependence on the total number of updates at each field value,
\emph{i.e.} the sweep rate, like in Fig.~\ref{figupdates}. The curves with
opposite ratios of spin to bond updates are similar only if the sum of the
updates is the same. The thin full line has the same ratio of the two as
the dash-dotted line, but shows the effect of faster cycling, with a more
pronounced figure eight, and noisier magnetization. (Note that the $N=10$
curve in Fig.~\ref{figupdates} has the same number of updates as the thick
full line in Fig.~\ref{figratio}.)

In fact, it turns out that the $\pm J$-Edwards-Anderson model is
extremely sensitive to the introduction of bond updates. Take the
$T/J=0.3$ curve in Fig.~\ref{figtemp}, obtained with a 1:1 ratio of bond
to spin updates. Going down to 1:200 barely changes it: it becomes
something like the $T/J=0.5$ curve in the same figure. But turning off
bond updates entirely collapses the main loop to the two boxes shown in
Fig.~\ref{figmain}.

Finally, we find that the figure-eight is quite sensitive to noise in the
bond couplings. A Gaussian spread with $\sigma=0.05J$ centered at $\pm J$ is
sufficient to flatten it almost completely. Thus while it can survive some
noise in the bond distribution --- \emph{i.e.} is not a singular effect ---
the figure-eight anomaly should at present be regarded as specific to the
$\pm J$ model.

\section{Experimental evidence}

Experimental observations of the figure-eight anomaly are not very rare,
but it is rare that anyone comments on them; in fact we found only one
paper in which this was explicitly mentioned as an unexplained
phenomenon~\cite{dosSantos95}. This may partly be because such forms can
be dismissed as artefacts, due to the integration of a noisy differential
signal $dM/dH$ in the saturation region, where the derivative is small.
Another possible reason is that they are sometimes observed~\cite{Kim03}
along with a number of other `misfeatures,' such as a marked absence of
inversion symmetry relative to the origin in the $(B,M)$ plane, so it is
not clear that they should be singled out for attention.

In Ref.~\cite{Kim03}, annealed amorphous samples of
Co$_{66}$Fe$_4$B$_{15}$Si$_{15}$ with a recrystallized surface were studied
by the magneto-optical effect. The surface was chemically etched to expose
the amorphous interior of annealed samples, and the changes in the hysteresis
curves recorded. After etching for ten minutes, the magnetic response became
softer and a thin but distinct figure-eight appeared in the saturation
region. Since the laser is a local probe, it was also possible to obtain
hysteresis curves at different physical positions on the as-quenched sample
surface. Some of these also show a figure eight, which the authors do not
comment. The width of the measured `boxy' main loop (large avalanche!) in the
field was of the same order as the range of the figure eight, just as found
in the model calculations here (see Fig.~\ref{figmain}).

In Ref.~\cite{Chesnel02}, a thin-film grating was subject to a perodic field
and the magnetic order observed by X-ray scattering. A perfectly reproducible
AF ordering was observed at both values of the coercive field, i.e. as the
sample passed through zero magnetization. This does not mean that the
positions of the effective AF (dipolar) couplings among the domains has been
changing, in fact quite the opposite, it is proof that it is fixed. The
interesting observation from our point of view occurred during sample
preparation. Direct observation of magnetic domains in `fresh' samples showed
they did not follow the lines of the grating, but were in fact quite random,
so that the grating geometry could barely be discerned. A careful
demagnetization process, with a slowly vanishing field, was necessary to
arrange the domains along the grating lines. During this process, the
effective couplings were evidently rearranged in space. Once they `fell into
line,' they did not evolve any further. We suggest a similar evolution is
happening in ordinary `fresh' samples, without control of domain positions.
First there is some rearrangement of domains under field cycling, amounting
to a spatial evolution (diffusion) of AF bonds. As the sample ages, the
domain positions should become fixed, pinned perhaps by impurities, and the
figure-eight anomaly should disappear. This scenario is corroborated in
Ref.~\cite{Kim03}, whose authors observed a hardening of the magnetic
response and disappearance of the anomaly after eight hours of annealing,
which recrystallized the surface. When the sample was etched as described
above, the anomaly reappeared.

\section{Discussion}

Hysteretic effects are to be expected whenever the approach to thermal
equilibrium is somehow hindered. One widely studied way in which this can
happen is when the equilibrium state is itself disordered, and the system
takes a long time to find this state among the many suboptimal, similarly
disordered ones. This is the case of spin glasses, especially in two
dimensions where the spin-glass transition temperature is zero. A different
way to hinder equilibration is provided by kinetic glasses, where the
underlying equilibrium state is ordered, but is never reached because of
kinetic obstruction.

The use of the disordered Ising model with fixed bond distribution to study
hysteresis assumes that the metastability involved in magnetic hysteresis may
have something in common with the metastability of spin glasses. This is
tacitly assumed in all such investigations of hysteresis with fixed bond
distributions. The present work investigates the alternative possibility,
that the delayed thermalization in magnetic hysteresis may be more like the
metastability of kinetic glasses. The figure-eight anomaly was identified as
a distinct and sensitive signal of this latter scenario. Experimental
observations~\cite{Chesnel02} allow that a real-space redistribution of
effective AF interactions between magnetic domains can be driven by an
external field, with response times within the experimental window. One would
intuitively expect this rearrangement capability to decrease as the sample
ages. Indeed this seems to be the case in the experiment mentioned
above~\cite{Kim03}.

The above observations open some interesting theoretical issues, as well. One
is a possible connection between the hidden order-disorder transition in the
bond background at the main avalanche, and the appearance of the figure-eight
anomaly in the state disordered by the field. A related issue is to
understand the origin of the power-law relaxation of the magnetization as a
function of the field. Another is the dependence of the anomaly on the
distribution of bond couplings --- we know, for example, that there is no
anomaly for a unimodular Gaussian distribution. These questions are currently
under investigation.

To conclude, we have given a simple interpretation of the figure-eight
anomaly, sometimes observed in magnetic thin films and tapes in the
saturation region, when the coercive fields involved are relatively weak,
of the order of 100~Oe. We believe it is related to a specific transient
process, which may be represented as a spatial rearrangement of AF
(dipolar) couplings in the sample, driven by the field.

\section{Acknowledgements}

Conversations with K.~Uzelac and I.~Campbell are gratefully acknowledged.
This work was supported by the Croatian Government under Project~$0119256$.

%\bibliography{/home/dks/vega/admin/fizika/clanci/htc}
%\bibliographystyle{apsrev}

\end{document}